\documentclass[12pt]{iopart}

\usepackage{iopams, amssymb, amsfonts, graphicx,psfrag,epsfig,color}
\newcommand{\alb}[1]{\textcolor{red}{#1}}
\newcommand{\albb}[1]{\textcolor{blue}{#1}}
\newcommand{\erf}{\mathrm{erf}}

\newcommand{\ii}{\mathsf{i}}
\newcommand{\mf}{f_0}

\newcommand{\en}{\mathcal{N}}
\newcommand{\E}{\mathrm{e}}
\newcommand{\D}{\mathrm{d}}
\newcommand{\Po}{P_{\mathrm{out}}}
\newcommand{\Pin}{P_{\mathrm{in}}}

\newcommand{\op}{f_+}
\newcommand{\om}{f_-}
\newcommand{\vp}{v_{\theta}}
\newcommand{\Is}{I^{<}}
\newcommand{\Il}{I^{>}}
\newcommand{\Ps}{P^{<}}
\newcommand{\Pl}{P^{>}}
\newcommand{\derpart}[2]{\frac{\partial #1}{\partial #2}}

\newcommand{\average}[1]{\left<{#1}\right>}
\newcommand{\p}[1]{\left({#1}\right)}
\newcommand{\pq}[1]{\left[{#1}\right]}
\newcommand{\pg}[1]{\left\{{#1}\right\}}

\usepackage{comment}
\excludecomment{mycomments}

\begin{document}

\title[]{Stochastic thermodynamics in  many--particle systems}

\author{ Alberto Imparato}

\address{Department of Physics and Astronomy, University of Aarhus, 8000 Aarhus C, Denmark}

\ead{imparato@phys.au.dk}

\begin{abstract}
We study the thermodynamic properties of a microscopic model of coupled oscillators that exhibits a dynamical phase transition from a desynchronized to a synchronized phase. 
We consider two different configurations for the thermodynamic forces applied on the oscillators, one resembling the macroscopic power grids, and one resembling autonomous molecular motors.
We characterize the input and the output power as well as the efficiency at maximum power, providing analytic expressions for such quantities near the critical coupling strength.
We discuss the role of the quenched disorder in the thermodynamic force distributions and show that such a disorder may lead to an enhancement of the efficiency at maximum power.
\end{abstract}

\vspace{2pc}

\maketitle

\section{Introduction}
The stochastic thermodynamics of microscopic systems has been the subject of intense investigation in recent years \cite{Seifert2012} in an attempt to extend basic concepts of macroscopic classical thermodynamics to the microscopic realm in general, and to out-of-equilibrium microscopic systems in particular. 
Notably a lot of effort has been devoted to the characterization of the efficiency of microscopic devices, that can transform heat or chemical energy into mechanical work.
While for the first type of devices (autonomous heat engines) the efficiency is bounded by the Carnot limit \cite{Esposito2009,Esposito2010}, in the case of isothermal engines the efficiency is constrained by the thermodynamic limit 1.
In both cases the upper limit is reached for quasi static operation, resulting in a vanishing power output.
Thus, a more relevant quantity to study is the efficiency at maximum power (EMP), that 
exhibits an interesting universal behaviour for different types of devices \cite{Esposito2009,Esposito2010, Esposito2009a,Seifert2011a,Golubeva2012,VandenBroeck2012,Golubeva2012a}. In particular the EMP in the linear regime is 1/2 of the 
maximal allowed value, while the behaviour beyond the linear regime depends on the details of the coupling between the energy producing and the energy consuming cycles.

Many of the theoretical studies have been directed toward the  characterization of the EMP in single devices such as soft nanomachines \cite{Seifert2011a}, single molecular motors \cite{Golubeva2012,VandenBroeck2012}, devices involving single electron
transport \cite{Esposito2009a,Sanchez2011,Esposito2012}, or single entropy-driven motors \cite{Golubeva2013a}.
However, an important class of microscopic devices  is represented by cell molecular motors, which  operate in crowded environments where their mutual interaction can become significant.  For example,
many kinesin motors walk on the same microtubule leading to traffic jam formation in some case \cite{Leduc2012}, while several motors can pull the same cargo  resulting in a strong cooperative effect \cite{Klumpp2005,Campas2006, Mueller2008, Holzbaur2010,Guerin2010}.
Furthermore, recent studies on synthetic nanomotors have been
conducted with the aim of reproducing the performance of their biological counterparts \cite{Kay2007,Liu2009,Lund2010}.
In this regard, it is important to note that it is now possible to engineer molecular spiders
that exhibit directional movement and behave like robots by carrying out a sequence of predetermined
actions \cite{Lund2010}. These artificial motors might in the future be organized in {\it teams}, to optimize, for example, their
transport properties and efficiency \cite{Rank2013}.
Thus it is clear that future research on  thermodynamic property optimization will deal with {\it teams} of interacting motors, where the dynamical phase these motors operate in becomes relevant.

One of the most studied models of interacting particles in out-of-equilibrium physics is the exclusion process, which exhibits three distinct dynamical phases with different densities and particle currents \cite{ASEP}. Furthermore, the exclusion process is often used to model molecular motors moving on a lattice, see, e.g. \cite{Lipowsky2001, Kolomeisky2007}. We have previously investigated the issue of EMP in isothermal interacting motors, modelled as an exclusion process on a single lattice \cite{Golubeva2012a,Golubeva2013}, or on a network \cite{Golubeva2014}.
In these studies, we found an  increase of the EMP in a many-motor system with respect to the single motor case, for a suitable choice of the model parameters. Remarkably, in \cite{Golubeva2012a,Golubeva2013} we found that the enhancement of the EMP occurs in a range of parameter values compatible  with the biological estimates for the molecular motor Kinesin.
From those studies we concluded that after a dynamical phase transition the dynamical response of the system to an external drive can change, leading in turn to a change in the thermodynamic properties. Specifically the dependence  of the delivered power on the driving thermodynamic forces may vary.

One of the main limitation that one faces when studying the thermodynamic properties of exclusion processes on a lattice, is that 
the intensity of interaction between the {\it motors}  can only be indirectly tuned by changing the kinetic parameters and thus the density of motors on the lattice \cite{Golubeva2012a,Golubeva2013,Golubeva2014}.
Furthermore, if one wants to study the effect of force disorder on the motor particles, one has to resort to numerical simulations, as no exact result exists for the exclusion process with heterogeneous  particles.

\begin{mycomments}

\alb{Say something on cyclic motors, why you use a periodic coordinate, there is no internal cycle for the molecular motors (both forces) all cycles are coupled to all}
\end{mycomments}

Instead, here we consider a  model of $N$ interacting microscopic particles, where the particle-particle interaction is an explicit  parameter, that can be tuned in order to drive a dynamical phase transition, from a weakly interacting -- incoherent system to a strongly interacting -- coherent system.
This model was originally introduced by  Sakaguchi in \cite{Sakaguchi88} as an extension of the Kuramoto model (KM) \cite{Kuramoto}, to study the synchronization of a group of interacting oscillators in contact with a reservoir at constant temperature.
Furthermore, the effect of quenched disorder in the thermodynamic force distribution can be taken into account within the present model.
The model is introduced and discussed in section \ref{sec:saka}.
The $N$ interacting particles can be viewed as a network of energy producers and users or as a system of interacting autonomous motors under the effect of thermodynamic forces.
Since the dynamical phase diagram can be obtained in terms of the particle interaction strength, temperature and force distribution, in section \ref{sec:stocterm} we will discuss how to calculate the relevant thermodynamic quantities, namely the deliverer and input power, and the efficiency.
We will consider two possible scenarios as far as the force distribution is concerned. In the first one, sec.~\ref{sec:either}, either a positive or a negative force is applied on each particle. In the second scenario, sec.~\ref{sec:both}, both a positive and a negative force is applied on the same particle.
We will thus discuss how to optimize the delivered power for the different types of network models, hence obtaining the EMP in terms of the interaction
intensity, and thus of the coherence between the particles' motion. 
We will finally discuss the effect of the quenched disorder in the force distribution on the thermodynamic quantities.

Interestingly, the model that we use here is a microscopic version of a model used to mimic macroscopic power grids. 
Indeed the dynamics of interconnected power grids can be  mathematically represented by a complex network of coupled oscillators \cite{Chertkov2013,Bullo13}, while at the macroscopic level one faces optimization problems different from the microscopic case, as shortly discussed in section \ref{macro:sub}.

\section{The Sakaguchi model}\label{sec:saka}

We consider a system of $N$ coupled oscillators, originally introduced by Sakaguchi \cite{Sakaguchi88}, described by the Langevin equation
\begin{equation}
\dot \phi_i(t)=f_i - \frac {K}{N} \sum_j \sin(\phi_i(t)-\phi_j(t))+ \eta_i(t),
\label{lang1}
\end{equation} 
where $f_i$ is an external constant force, and the Gaussian noise $\eta_i$ obeys the fluctuation--dissipation relation
\begin{equation}
\average{\eta_i(t) \eta_j(t')}=2 k_B T \delta_{ij} \delta(t-t').
\label{FD:ed}
\end{equation} 
Notice that we have chosen the system units such that the external force $f_i$ has dimension of frequency, which corresponds to taking the friction coefficient in eq.~(\ref{FD:ed}) equal to one.

By introducing the complex order parameter 
\begin{equation}
\sigma(t) \exp(\ii \psi(t)) =\frac 1 N \sum_j  \exp(\ii \phi_j(t)),
\label{ordpar}
\end{equation} 
where $0\le \sigma(t)\le 1$ measures the system coherence and $\psi(t)$ is the common average phase, 
equation (\ref{lang1}) becomes
\begin{equation}
\dot \phi_i=f_i - K \sigma \sin(\phi_i-\psi)+ \eta_i,
\label{lang2}
\end{equation} 
where we understood the dependence on time.
Let $\mf$ be the mean deterministic force, calculated  over the $N$ oscillator sample $\mf=\sum_j f_j/N$, we expect that  the center of mass will oscillate with the frequency $\mf$, so we can set $\psi(t)=\mf t +\psi_0$ and thus 
we can redefine the dynamical variables as $\theta_i=\phi_i-\psi(t)$, so as eq.~(\ref{lang2}) reads
\begin{equation}
\dot \theta_i=\omega_i - K \sigma \sin(\theta_i)+ \eta_i,
\label{lang3}
\end{equation} 
where we have redefined the external force as $\omega_i=f_i -\mf$.

\begin{mycomments}
{\it 
 Is it correct to take $\psi(t)=\mf t +\psi_0$ at this stage, or should I rather take $\psi(t)=v t +\psi_0$ and then use $v$ as a free variable, together with $\sigma$, to solve eq.~(\ref{eqsig}) 
}
\end{mycomments}

In principle eq.~(\ref{lang3}) represents a set of $N$ coupled equations for the variables $\theta_i$, since $\sigma$ is given by eq.~(\ref{ordpar}). However, as $N\to \infty$, one can replace the actual value of $\sigma$ with its mean field value, and so  eq.~(\ref{lang3}) becomes uncoupled. Such a mean field value can be obtained self consistently as discussed below.
Eq.~(\ref{lang3}) corresponds to a Brownian particle moving in a periodic potential under the effect of a constant drift force $\omega_i$. 
Here and in the following we assume that the system reaches a  steady state in the long time limit. In the course of this paper, we will discuss this assumption where relevant. 
The Langevin equation can be reformulated in terms of a Fokker-Planck (FP) equation for the probability distribution function (PDF) of finding the particle $i$ at position $\theta$ at time $t$
\begin{equation}
\partial_t p(\theta, \omega_i, t)=\partial_\theta\pq{ (K \sigma \sin \theta -\omega_i) p +T \partial_\theta p}.
\label{FP:eq}
\end{equation} 
Thus, the stationary probability distribution function (PDF) of the position of such a particle reads \cite{Golubeva2012,vanKampen1981}
\begin{equation}
p(\theta,\omega_i)=\en \beta  \E^{-\beta (K \sigma \cos \theta + \omega_i \theta) }\left[\frac{I(2 \pi)}{1-\exp\left(-\beta 2 \pi \omega_i \right)}-I(\theta)\right],
\label{pss}
\end{equation} 
where $I(x)=\int_0^x \D y \exp\left[\beta (K \sigma \cos y + \omega_i y) \right]$, and $\en$ is a normalization constant depending implicitly on $\beta=1/T$, $K\cdot \sigma$ and $\omega_i$.
The steady--state probability current thus reads $J_{ss}=\en$, and the particle steady--state velocity reads
\begin{eqnarray}
v_\theta(K \sigma, \omega_i, T)= 2 \pi \en && \nonumber \\
= 2 \pi  \pg{ \beta \int_0^{2 \pi} \D \theta\,  \E^{-\beta (K \sigma \cos \theta + \omega_i \theta) }\left[\frac{I(2 \pi)}{1-\exp\left(-\beta 2 \pi \omega_i \right)}-I(\theta)\right]}^{-1}&&.
\label{v:ss}
\end{eqnarray} 

As $N\rightarrow \infty$, we can adopt a continuous description, where the constant forces acting on the oscillators are distributed according to the probability distribution $g(f)$ with mean value $\mf$. By introducing the shifted force distribution 
\begin{equation}
g_0(\omega)=g(\mf+\omega),
\end{equation} 
the self-consistent equation for the modulus $\sigma$ of the complex order parameter, characterizing the degree of order or coherence in the configuration of the variables $\theta_i$ is then given by \cite{Sakaguchi88}
\begin{equation}
\sigma \E^{\ii \psi_0}=\int \D \omega\,  g_0(\omega) \int_0^{2 \pi} \D  \theta\, p(\theta,\omega) \exp(\ii \theta)
\label{eqsig}
\end{equation} 
which can be decomposed into its real and imaginary part
\begin{eqnarray}
\sigma \cos(\psi_0)&=&\int  \D \omega\,  g_0(\omega) \int_0^{2 \pi} \D  \theta\, p(\theta,\omega) \cos \theta\, ,\\
\label{eqcos}
\sigma \sin(\psi_0)&=&\int  \D \omega\,  g_0(\omega)  \int_0^{2 \pi} \D  \theta\, p(\theta,\omega) \sin \theta\, .
\label{eqsin}
\end{eqnarray} 
\begin{mycomments}
{\it Notice the different approach of page 39, eq 4.8 of \cite{Strogatz89}}
\end{mycomments}

By assuming that the force distribution $g(f)$ is symmetric around $\mf$, 
and noticing that $p(\theta,-\omega)=p(-\theta, \omega)$, the imaginary part on the right--hand side of eq.(\ref{eqsig}) vanishes, and so one is left with
 \begin{equation}
\sigma =\int \D \omega\,  g_0(\omega) \int_0^{2 \pi} \D  \theta\, p(\theta,\omega) \cos(\theta),
\label{eqsig1}
\end{equation} 
 whose solution provides the mean field value for $\sigma$.

\begin{mycomments}
{\it How do you prove that the imaginary part on the right--hand side of eq.(\ref{eqsig}) vanishes? Consider that in eq.~(\ref{lang3}) when you take $\omega\rightarrow-\omega$ (with $\omega>0$), the landscape is tilted in the negative direction. So  $p(\theta,-\omega)=p(-\theta, \omega)$. Now, when $g_0(\omega)$ is symmetric, we have $\int \D \omega\,  g_0(\omega) \int_0^{2 \pi} \D  \theta\, p(\theta,\omega) \exp(\ii \theta)=\int_0^{+\infty}  \D \omega\,  g_0(\omega) \int_0^{2 \pi} \D  \theta\, (p(\theta,\omega)+p(-\theta,\omega))  \exp(\ii \theta)$ and $ (p(\theta,\omega)+p(-\theta,\omega))$ is an even function of $\theta$, so its projection onto $\sin(\theta)$ gives zero.}
\end{mycomments}

As discussed in \cite{Sakaguchi88}, for $N\rightarrow \infty$ this model exhibits a critical coupling strength $K_c$, such that for $K>K_c$ the systems exhibits a dynamical phase transition with synchronization $\sigma>0$, while the system is incoherent for $K<K_c$, and each particle described by the coordinate $\theta_i$ oscillates with its proper frequency $\omega_i$.
Thus, for $K\gtrsim K_c$ we expect $\sigma$ to be positive but small, and we can expand eq.~(\ref{eqsig1})  in powers of $\epsilon=K \sigma/T$, obtaining
\begin{eqnarray}
\sigma&=&\frac{ K \sigma T}{2} \int_{-\infty}^{+\infty} \D \omega\,  \frac{g_0(\omega)}{\left(T^2+\omega ^2\right)} \left[1-\frac{K^2 \sigma ^2 \left(T^2-2 \omega ^2\right)}{2 \left(T^2+\omega ^2\right) \left(4 T^2+\omega ^2\right)}\right. \nonumber \\
 &&\qquad \left.+ \frac{ K^4 \sigma^4 \left(3 T^4-17 T^2 \omega ^2+4 \omega ^4\right)}{4 \left(T^2+\omega ^2\right)^2 \left(4 T^2+\omega ^2\right) \left(9 T^2+\omega ^2\right)}\right]
 + O\p{ \epsilon^7}, 
\label{eqcos1}
\end{eqnarray} 
while expanding eq.~(\ref{v:ss}) the average velocity of the dynamical variable $\theta$ reads
\begin{eqnarray}
v_{\theta}(\omega)&=& \omega  \left[ 1-\frac{ K^2 \sigma ^2}{2(T^2+\omega ^2)}+\frac{K^4 \sigma ^4 \left(5 T^2-\omega ^2\right)}{8 \left(T^2+\omega ^2\right)^2 \left(4 T^2+\omega ^2\right)}\right. \nonumber \\ 
&&\quad \left.- \frac{K^6 \sigma ^6 \left(23 T^4-24 T^2 \omega ^2+\omega ^4\right)}{16 \left(T^2+\omega ^2\right)^3 \left(4 T^2+\omega ^2\right) \left(9 T^2+\omega ^2\right)} \right]
 +O\p{ \epsilon^8}.
\label{v_exp}
\end{eqnarray} 
It is worth to note that the first two coefficients of the expansion (\ref{eqcos1}) were recently re-derived in ref.~\cite{Sasa15} by using a different approach: the author  mapped the deterministic time evolution of the Kuramoto model order parameter into a stochastic process as given by eq.~(\ref{lang1}), and applied a fluctuation relation to the thermodynamic irreversible work  done on such a stochastic system.

Inspection of eq.~(\ref{eqcos1}) provides the critical coupling strength for which a non-vanishing solution to that equation appears 
\begin{equation}
K_c=2 \pq{\int_{-\infty}^{+\infty} \D \omega \,  g_0(\omega) \frac{T}{ \left(T^2+\omega ^2\right)}}^{-1}. 
\label{kc:def}
\end{equation} 

The value of the order parameter $\sigma$ as a function of $K$ and $T$, for $K>K_c$  can be obtained by solving eq.~(\ref{eqcos1}), which gives  
\begin{mycomments}
{\it (fourth\_order.nb)}
\end{mycomments}
\begin{eqnarray}
\sigma&=&\pq{ \frac{I_3-\sqrt{I_3^2+4 I_5 \left(\frac{K_c-K}{K K_c}\right)}}{2 I_5  K^2}}^{1/2}\nonumber \\
&\simeq& \sqrt{\frac{K-K_c}{K_c^4  I_3}}\pq{1+ \frac{(K-K_c) (I_5 -3 I_3 ^2 K_c) }{2 I_3^2 K_c^2}}  + O \p{\Delta K^{5/2}},\label{sigma_exp}
\end{eqnarray} 
where we have introduced the quantity $\Delta K=K-K_c$, and 
\begin{eqnarray}
I_3&=& \int_{-\infty}^{+\infty} \D \omega\,   \frac{g_0(\omega) T \left(T^2-2 h ^2\right)}{4 \left(T^2+h ^2\right)^2 \left(4 T^2+h ^2\right)},\\
I_5&=&\int_{-\infty}^{+\infty} \D \omega\, g_0(\omega)\frac{  T \left(3 T^4-17 T^2 \omega ^2+4 \omega ^4\right)}{8 \left(T^2+\omega ^2\right)^3 \left(4 T^2+\omega ^2\right) \left(9 T^2+\omega ^2\right)}.
\end{eqnarray} 

The figures contained in this manuscript and discussed in the following, are made by using the approximate expressions (\ref{v_exp}) and (\ref{sigma_exp}), and choosing the values of the parameters such that $\sigma \lesssim0.5$.

A few comments on the velocity $v_\theta(\omega)$ are now in order. Such a quantity represents the velocity of the dynamical variable $\theta$ and is thus the velocity deviation of a particle under the effect of a force $\omega$ with respect to the center of mass velocity $f_0$. Inspection of eq.~(\ref{lang3}) suggests that for $K<K_c$ (thus for $\sigma=0$)   $v_\theta(\omega)=\omega$. On the other hand  $v_\theta(\omega)$ goes to zero as $K$ increases above $K_c$: the higher $K$ the higher are the barriers of the periodic force in eq.~(\ref{lang3}), while  $\sigma$ is also an increasing function of $K$.

\section{Macroscopic power grids} \label{macro:sub}
Here we want to establish contact between the network of microscopic oscillators discussed in the previous section, and a corresponding macroscopic model used to describe macroscopic power grids. This analogy will be useful in the next section, where we will introduce the thermodynamic forces and powers characterizing the microscopic model.
Given that in the macroscopic realm, one deals with alternating current (AC) networks, 
in terms of equations, this amounts to write a system of equations for the phase angle $\phi_i$ for  both the generators and the users \cite{Bullo13}
which correspond to an extended KM 
\begin{eqnarray}
M_i \ddot \phi_i +D_i \dot \phi_i&=&\omega_i -\sum_{j=1}^N a_{ij} \sin(\phi_i-\phi_j); \qquad \mathrm{if}\, i\, \mathrm{is\, a\, generator}\label{pow:grid1}\\
D_i \dot \phi_i&=&\omega_i -\sum_{j=1}^N a_{ij} \sin(\phi_i-\phi_j), \qquad \mathrm{if}\, i\, \mathrm{is\, a\, consumer}; \label{pow:grid2}
\end{eqnarray} 
with inertia coefficients $M_i$ (representing, e.g., the large rotational inertia in turbine generators), viscous damping $D_i$, power injection $\omega_i$ (consumption if $\omega_i<0$) and power flows along the lines  $a_{ij} \sin(\phi_i-\phi_j)$ with coupling strength $a_{ij}$.
This model exhibits a synchronized phase that corresponds to a power grid that operates in a steady state with spatially uniform frequency.

In this kind of power network one may want to determine the optimal  operation conditions to, e.g., avoid blackouts.
For example in  a static grid with a fixed number of producers and users, where the generators have a maximum power capability, and the users are characterized by
a well known average consumption, one may want to optimize the grid in order to avoid that the consumers' load exceed the generators
capability, leading to blackouts. This can be done by using
optimization algorithms for graphs, see, e.g., \cite{Chertkov2009}.

Another, perhaps more interesting optimization problem considers a power grid as a dynamical system, where
both energy producers and users can be dynamically connected or disconnected over a given time
period. On the producers’ side, this is the case of renewable energy sources, such as wind and solar
power, which are stochastic in nature and often uncontrollable, resulting in severe difficulties in the
maintenance of the balance between load and generation \cite{He2011}. Thus in the future opportunistic
users can access the energy system according to the availability of system resources and differently
from the "always-on'' demand of traditional energy users, their consumption can exhibit peaks of activity. In this scenario, the challenge is  to coordinate and manage dynamically interacting power grid participants.

\section{Stochastic Thermodynamics of the microscopic model} \label{sec:stocterm}
In section \ref{sec:saka} we have discussed  the dynamical properties of the model system  we will use in the present paper. We can now turn our attention to its thermodynamic properties, namely the input and delivered power, and the system efficiency as a global motor.

One can consider two possible scenarios as far as the forces applied on each single particle are concerned.

In the first case 
the forces acting on the motor system can be either positive or negative, 
thus resembling the macroscopic power grids of power plants and consumers considered, e.g., in \cite{Filatrella08,Rohden12, Bullo13,Olmi14}.
Differently from those works, we consider here microscopic oscillators, in the over-damped regime, and with white noise acting on them.
In this scenario, taking inspiration from the macroscopic realm,  one may call {\it users} those oscillators with a negative force acting on them $f_i<0$, and {\it producers} those oscillators with a positive force $f_i>0$, and a single force distribution characterizes the system.

The second possible scenario resembles the case of {\it molecular motors}, where both a negative ($f_i^-<0$) and a positive force ($f_i^+>0$) are applied on the same particle $i$. This is the case in, e.g., biological molecular motors such as kinesin and myosin \cite{Alberts2007, Howard2001} where the energy extracted by ATP hydrolysis drives the motor forward (corresponding to $f_i^+>0$) while the motor does work to carry a cargo, modelled by a negative load (corresponding to $f_i^-<0$)
In this case one deals with two different distributions of forces, $g_+(f_+)$ and  $g_-(f_-)$.
The homogeneous case, where the {\it same} positive $f_+$ and negative force $f_-$ where applied on all the motors, modelled as diffusing particle on a lattice with an exclusion rule, was studied, e.g., in \cite{Golubeva2012a,Golubeva2013,Golubeva2014}.

In both scenarios, in order for the system to behave globally as a motor, and to perform work against the negative forces, we must require the center of mass to have an average positive velocity, and thus $\mf>0$.

In the following, we will consider, for both scenarios, the delivered power $\Po$ and the input power $\Pin$, and optimize $\Po$ wrt different parameters. We will thus characterize the efficiency at maximum power (EMP) $\eta^*=\Po^*/\Pin^*$ mainly in proximity of the dynamical phase transition, and where possible, we will discuss the thermodynamic properties of the system in the whole range of parameter space.

\subsection{Single force distribution}\label{sec:either}

We consider here the stochastic thermodynamics of a system with either negative or positive forces applied on each oscillators, and distributed according to the single PDF $g(f)$.

We can thus introduce the relevant thermodynamic quantities, namely the average 
input power, absorbed by the producers, and the average output power released by the users. Recalling that $v_\theta$ as given by eq.~(\ref{v:ss}) gives the deviation of the $i$-th particle's average velocity from the center of mass velocity $f_0$, the average output and input power read
\begin{eqnarray}
\Po&=&-\int_{-\infty}^{0} \D f\,  g(f) \pq{v_\theta(f-\mf)+\mf}f\nonumber \\
&=&-\int_{-\infty}^{-\mf} \D \omega\,  g_0(\omega) \pq{v_\theta(\omega)+\mf}(\omega+\mf)\label{po:def}\, ,\\
\Pin&=&\int^{+\infty}_{0} \D f\,  g(f) \pq{v_\theta(f-\mf)+\mf}f\nonumber \\
&=&\int^{+\infty}_{-\mf} \D \omega\, g_0(\omega)\pq{v_\theta(\omega)+\mf}(\omega+\mf)\, .\label{pi:def} 
\end{eqnarray} 
while the thermodynamic efficiency of the system reads
\begin{equation}
\eta=\frac{\Po}{\Pin}.
\end{equation} 

Substituting eqs.~(\ref{v_exp}) and (\ref{sigma_exp}) into (\ref{po:def}) and (\ref{pi:def}) the output and input power becomes, up to the second order in $K-K_c$, 
\begin{eqnarray}
\Po&=&\Ps_0+\frac{K-K_c}{K_c^2  I_3} \Is_2 +  \frac{(K-K_c)^2 \left(\Is_2 I_5+I_3 \Is_4 -\Is_2 I_3^2 K_c\right)}{I_3^3 K_c^4}  , \label{ps:exp} \\
\Pin&=&\Pl_0+\frac{K-K_c}{K_c^2  I_3} \Il_2 + \frac{(K-K_c)^2 \left(\Il_2 I_5+I_3 \Il_4 -\Il_2 I_3^2 K_c\right)}{I_3^3 K_c^4}\, , \label{pl:exp} 
\end{eqnarray} 
where 
\begin{eqnarray}
\Ps_0&=& -\int_{-\infty}^{-\mf} \D \omega g_0(\omega)(\omega+\mf)^2 <0 \label{ps0} , \\
\Is_2&=& \int_{-\infty}^{-\mf} \D \omega g_0(\omega)\frac{(\omega+\mf) \omega}{2(T^2+\omega ^2)} \ge0, \\
\Is_4&=& -\int_{-\infty}^{-\mf} \D \omega g_0(\omega)\frac{(\omega+\mf) \omega \left(5 T^2-\omega ^2\right)}{8 \left(T^2+\omega ^2\right)^2 \left(4 T^2+\omega ^2\right)},  
\end{eqnarray} 
with analogous definitions for $\Pl_0$, $\Il_2$ and $\Il_4$.
We notice that, in absence of partial synchronization ($K<K_c$, $\sigma=0$), i.e., when the users and the producers are decoupled, eq.~(\ref{po:def}), (\ref{ps:exp}), and (\ref{ps0})  predict that the delivered power is negative. Since $v_\theta(\omega)=\omega$  for $K<K_c$, the {\it users} oscillates with their proper frequency (force) which is negative, and so the product of the applied forces times the average velocity is positive.
The term $v_\theta(\omega)$ in eq.~(\ref{po:def}) is always negative, as the integration variable runs over negative value. Thus by increasing $K$ above $K_c$ the modulus of $v_\theta(\omega)$ decreases and tends to zero for very large $K$. This implies that for some value of $K$ the rhs of eq.~(\ref{po:def}) becomes positive, such values depending on $f_0$ and $T$, and on the details of the distribution $g_0(\omega)$, e.g. its width.
\subsubsection{Optimization}
Here we aim at optimizing the delivered power eq.~(\ref{po:def}) wrt some of the relevant parameters.

Optimization wrt the coupling strength $\partial \Po/\partial K=0$ gives:
\begin{equation}
\derpart {\Po}{K}=- \int^{-f_0}_{-\infty} \D \omega \,  g_0(\omega)\omega \partial_K v_{\theta}(K, \omega,T).
\label{derK:po}
\end{equation} 
Recalling  that the average velocity $v_{\theta}$  goes to zero   as $K$ increases above $K_c$, we find that, for $\omega<0$, $v_{\theta}(\omega,K)$ is an increasing function of $K$, ranging from $\omega$ for $K<K_c$ and approaching zero as $K\to \infty$.
Thus, from eq.~(\ref{derK:po}) it follows
\begin{equation}
\derpart {\Po}{K}=\cases{0, & $\mathrm{if}\, K<K_c,$\\
 \ge 0 & $\mathrm{if}\, K\ge K_c.$}
\end{equation} 
Similarly one finds 
\begin{equation}
\derpart {\Pin}{K}=\cases{0, & $\mathrm{if}\, K<K_c,$\\
 \le 0 & $\mathrm{if}\, K\ge K_c.$}
\end{equation} 

\begin{mycomments}
{\it 
Trick for that, write
\begin{equation}
\derpart {\Pin}{K}=\int_{-f_0}^{+\infty} \D \omega \, g_0(\omega) (\omega+f_0) \partial_K v_\theta=\int_{-f_0}^{+f_0}\dots+   \int_{f_0}^{+\infty}\dots
\end{equation} 
The second integral is negative for sure.
We then have for the first integral 
\begin{eqnarray}
&& \int_{-f_0}^{+f_0}  \D \omega \, g_0(\omega) (\omega+f_0) \partial_K v_\theta=\partial_K \int_{-f_0}^{+f_0}  \D \omega \, g_0(\omega) (\omega+f_0)  v_\theta\nonumber \\
&=& \partial_K \int_{-f_0}^{+f_0}  \D \omega \, g_0(\omega) \omega  v_\theta=\partial_K \int_{0}^{+f_0}  \D \omega \, g_0(\omega) \omega  v_\theta\nonumber\\
&=& \int_{0}^{+f_0}  \D \omega \, g_0(\omega) \omega  \partial_K v_\theta\le 0
\end{eqnarray} 
}
\end{mycomments}
Thus, if $K$ is the free parameter, the optimal delivered power is achieved for $K\to \infty$, corresponding to the limit of strong coupling between users and producers, with an EMP $\eta^*=\average{f_-}/\average{f_+}$ as obtained by eqs.~(\ref{po:def}) and (\ref{pi:def}), where $\average{f_-}$ and $\average{f_+}$ are the average negative and positive forces, respectively.

No similar inequalities can be found when one tries to maximize $\Po$ with respect to other parameters, for example $f_0$. So one should consider specific cases for the force distribution in order to study the relevant thermodynamic quantities.

\subsubsection{A specific distribution}
In order to exemplify the results discussed in this section, here we consider the 
specific distribution 
\begin{equation}
g(f)=\frac 1 2 \pq{\delta(f-(\mf+s))+ \delta(f-(\mf-s))},
\end{equation}
where $s^2$ is the variance of the distribution, with $s>\mf >0$, 
i.e., there are just two types of oscillator, the {\it users} with an applied force $\mf-s<0$ and the {\it producers} with an applied force   $\mf+s>0$.
The shifted force distribution thus reads
\begin{equation}
g_0(\omega)=\frac 1 2 \pq{\delta(\omega-s)+ \delta(\omega+s)}.
\label{g0s}
\end{equation}
For such a distribution the critical coupling strength reads
\begin{equation}
K_c=2\frac {(s^2 +T^2)} T.
\label{Kcs}
\end{equation} 
This corresponds to the bimodal distribution considered in \cite{Acebron05}, where the linear stability of the incoherent solution $p(\theta, \omega)=1/ 2 \pi$  of the FP equation~(\ref{FP:eq}) was studied, corresponding to the non-synchronized phase $\sigma=0$. The authors showed that in the limit $N\to \infty$ the FP equation eq.~(\ref{FP:eq}) exhibits a steady state solution for $s<T$ and $K>K_c$, while for $s>T$ and $K>4 T$ the system exhibits an oscillatory state.
In the following we will thus take $s<T$, and use the steady state solution (\ref{pss}) to eq.~(\ref{FP:eq}).
Equations (\ref{ps:exp})-(\ref{pl:exp}) thus become
\begin{eqnarray}
\Po&=&\frac 1 2 (s-f_0) ( f_0-v_\theta(s)) , \label{xx} \\
\Pin&=&\frac 1 2 (s+f_0) ( f_0+v_\theta(s)) , \label{yy} 
\end{eqnarray} 
We recall  that for $K<K_c$ (i.e. for $\sigma=0$), $v_\theta(s)=s$, and because of  the condition $s>f_0$ we have $\Po<0$, i.e. when  the producers and users are not coupled, the users oscillates with their proper frequency $f_0-s$ resulting in a negative $\Po$ . The delivered power  will become positive for some value of $K>K_c$, when $v_\theta(s)$ in eq.~(\ref{xx}) becomes smaller than $f_0$. 

\subsubsection{Optimization}
Here we optimize the delivered power for the force distribution  (\ref{g0s}), which corresponds to a system
 where on each oscillator there is either a positive $f_0+s$ or a negative $f_0-s$ force with probability 1/2.
From eq.~(\ref{v_exp}) we easily obtain the expression for the velocity deviation from the center of mass up to the fourth order in $\sigma$ 
\begin{eqnarray}
 v_\theta(s)&=&s \left[1-\frac{ K^2 \sigma ^2}{2(s^2+T^2)}+\frac{  K^4 \sigma ^4 \left(5 T^2-s^2\right)}{8\left(s^2+T^2\right)^2 \left(s^2+4 T^2\right)}\right. \nonumber \\ 
&& \quad \left.- \frac{K^6 \sigma ^6 \left(23 T^4-24 T^2 s ^2+s ^4\right)}{16 \left(T^2+s ^2\right)^3 \left(4 T^2+s ^2\right) \left(9 T^2+s ^2\right)} \right],
\label{v_s}
\end{eqnarray} 
while the order parameter, as given by eq.~(\ref{sigma_exp}), becomes
\begin{equation}
\sigma=\sqrt{\frac {\Delta K T (s^2+4 T^2)}{K_c^2(T^2-2 s^2)}}\pq{1+\Delta K \frac{8 s^6  + 97 s^4 T^2 - 40 s^2 T^4 + 15 T^6}{2  K_c (T^2-2 s^2 )^2  (s^2 + 9 T^2)}}, 
\label{s_s}
\end{equation} 
up to the order 3/2 in $\Delta K$.

\begin{mycomments}
{\it 
Use this argument
introduce $y=\sigma^2$, from eq.~(\ref{s_s}) we see that $y$ is of order $\Delta K^3$, while in eq.~(\ref{v_s}) we have terms of order $\Delta K$, $\Delta K^2$ and $\Delta K^3$.
}
\end{mycomments}

We can now optimize $\Po$, as given by eq.~(\ref{xx}),  wrt to different parameters:

$i)$ By optimizing wrt to $K$ at fixed $f_0$ and $s$: $\partial_K \Po=0$, one obtains
\begin{equation}
\derpart {\Po}{K}=-1/2 (s-f_0) \derpart {v_\theta(s)}{K}>0,
\end{equation} 
since $v_\theta(s)$ is a decreasing function of $K$, as already discussed above for a general force distribution $g_0(\omega)$.

$ii)$  By optimizing $\Po$ wrt the average force, at fixed $K$ and $s$, $\partial_{f_0} \Po=0$,
 \begin{mycomments}
{\it see two\_omega\_minus0.nb, bottom}
\end{mycomments}
 one obtains
\begin{equation}
f_0^*(s,K)=\frac{v_\theta(s)+s}{2}
\end{equation} 
and since the condition $s\ge v_\theta(s)>0$ holds for any $K>0$, we have $s>f_0^*(s,K)>s/2$.
\begin{figure}[h]
\center
\psfrag{s0.3}[lc][lc][.7]{$s=0.3$}
\psfrag{s0.2}[lc][lc][.7]{$s=0.2$}
\psfrag{s0.1}[lc][lc][.7]{$s=0.1$}
\psfrag{K}[ct][ct][1.]{$K$}
\psfrag{dK}[ct][ct][1.]{$(K-K_c)/K_c$}
\psfrag{etas}[ct][ct][1.]{$\eta^*$}
\includegraphics[width=8cm]{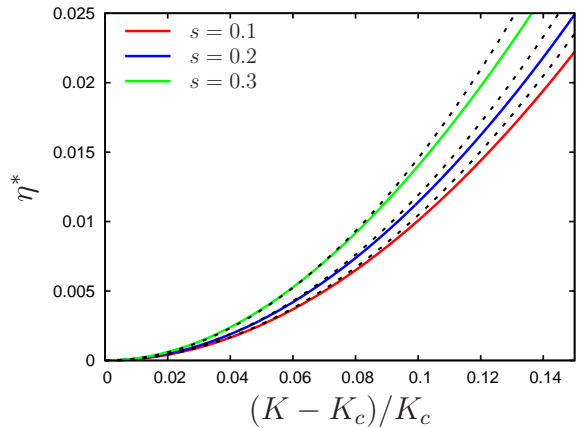}
\caption{EMP $\eta^*$ as obtained by maximizing $\Po$ wrt $f_0$,  as a function of $K$ for different values of the quenched disorder standard deviation $s$. Here $T=1$. The dashed lines correspond to the approximated expression (\ref{etass_app}). }
\label{fig:either}
\end{figure}

We can thus calculate the delivered and the input power,  and the efficiency at the maximum 
\begin{eqnarray}
\Po^*&=&\frac 1 8 (s-v_\theta(s))^2 \label{poss}\\
\Pin^*&=& \frac 1 8 (3s+v_\theta(s))(s+3 v_\theta(s))\label{pins}\\
\eta^*&=&\frac{(s-v_\theta(s))^2}{(3 s+v_\theta(s)) (s+3 v_\theta(s))}\simeq \frac{\Delta K^2 \left(s^2+4 T^2\right)^2}{16 K_c^2 \left(T^2-2 s^2\right)^2} \label{etass}\\
&\simeq& \frac {\Delta K^2}{T^2} \p{ \frac 1{4 }+\frac {5 s^2}{8 T^2} },\label{etass_app}
\end{eqnarray} 
where we have used (\ref{v_s}) and (\ref{s_s}) to expand $\eta^*$ up to the lowest order in $\Delta K$ and $s/T$. Plots of $\eta^*$ as a function of $K$ for different values of $s$ are shown in fig.~\ref{fig:either}.
Inspection of this figure, as well as of eqs.~(\ref{poss}), (\ref{pins}) and (\ref{etass}) suggests that, for fixed $s$, when $K$ increases above $K_c$, the optimal output power (\ref{poss}) increases, the optimal input power (\ref{pins}) decreases, and this results in an increase of the  EMP  (\ref{etass}). This is a consequence of the fact that $v_\theta(s)\to 0$ in the limit $K\to \infty$, where $\eta^*=1/3$.

Inspection of figure~\ref{fig:either}, as well as of eq.~(\ref{etass_app}), suggests that a higher degree of quenched disorder, as parametrized by $s$, leads to a larger EMP close to the dynamical phase transition.
However, the analysis of the behaviour of eqs.~(\ref{poss}), (\ref{pins}) and (\ref{etass}) at fixed $\Delta K$ and varying $s$ is not so straightforward.
Graphical analysis of eqs.~(\ref{poss}), (\ref{pins}) (not shown) indicates that both $\Po^*$ and $\Pin^*$ increase with $s$, with $\Po^*$ increasing faster. This graphical check can be done in the range of parameters where eqs.~(\ref{v_s}) and (\ref{s_s}) holds, i.e. close to the critical point. However, it is worth to note that for a fixed $\Delta K$, one finds $\Po^*(s=0)=0$.  Furthermore inspection of eq.~(\ref{lang3}) also suggests that  $v_\theta(s)\to s$, as $s\to\infty$, and being  $\Po^*$ a positive quantity, it must have at least one maximum for $s\in[0,+\infty[$. On the other hand, $\Pin^*$, as given by eq.~(\ref{pins}) in an increasing function of $s$. Accordingly $\eta^*$ has  at least one maximum for $s\in[0,+\infty[$.

\begin{mycomments}
\albb{ The argument for the last claim could be that if $s_0$ is the max. for $\Po^*$, then $\eta^*$ is decreasing for $s>s_0$ (write the first derivative of $\eta^*$ and check that it is negative), on the other hand, eq.~(\ref{etass_app}) shows that  $\eta^*$ increases for small $s$, so the first derivative changes sign for $s$ somewhere between 0 and $s_0$.

Note that I also need $\Delta K$ fixed here, but in principle $K_c$ depends on $s$, the larger $s$ the larger $K_c$.
}

\alb{Why?? It's a non trivial effect, both $\Po^*$ and $\Pin^*$ increase with $s$ but $\Po^*$ seems to increase more rapidly, as results from a graphical check.
Note also that from eq.(\ref{poss}), one has  $\Po^*(s=0)=0$ but also $\Po^*(s\to \infty)=0$ so there must be a maximum somewhere.}

{\it 
We can now investigate how the EMP varies with the free parameters $K$ and $s$.
By deriving eq.~(\ref{etass}) wrt $K$, at constant $s$,  
we have 
\begin{equation}
\derpart{\eta^*}{K}=\derpart{\eta^*}{v_\theta} \derpart {v_\theta(s)}{K}=-\frac{16 s (s^2-v_\theta^2(s))}{\pq{(3 s+v_\theta(s)) (s+3 v_\theta(s))}^2} \derpart {v_\theta(s)}{K}\ge0,
\end{equation} 
and so the EMP increases with $K$.
By deriving eq.~(\ref{etass}) wrt $s$, we obtain 
\begin{equation}
\derpart{\eta^*}{s}=\frac{16 (s^2-v_\theta^2(s))(v_\theta(s)-s v_\theta'(s))}{\pq{(3 s+v_\theta(s)) (s+3 v_\theta(s))}^2},
\end{equation} 
while no general statement can be done on the sign of this derivative, from eq.~(\ref{etass_app}), and considering that $K_c$ is given from eq.~(\ref{Kcs}), one finds that $\eta^*$ increases as $s$ decreases below (but close to) $\sqrt{K T /2 -T^2}$. The requirement $s<\sqrt{K T /2 -T^2}$ corresponds to $K>K_c$.
It is worth to note that also the maximum power $\Po^*$ (\ref{poss}) increases for increasing  $K>K_c$, as a consequence of the fact that $v_\theta(s)$, the deviation of the particle velocity from the center of mass velocity $f_0$, is a decreasing function of $K$, with $v_\theta(s)\to 0$ when $K\gg K_c$.
}
\end{mycomments}

\begin{figure}[h]
\center
\psfrag{etas}[ct][ct][.8]{$\eta^*$}
\psfrag{f0}[ct][ct][.8]{$f_0$}
\psfrag{K2.25}[rc][rc][.7]{$K=2.25$}
\psfrag{K2.5}[rc][rc][.7]{$K=2.5\, $ }
\psfrag{K2.75}[rc][rc][.7]{$K=2.75$}
\includegraphics[width=8cm]{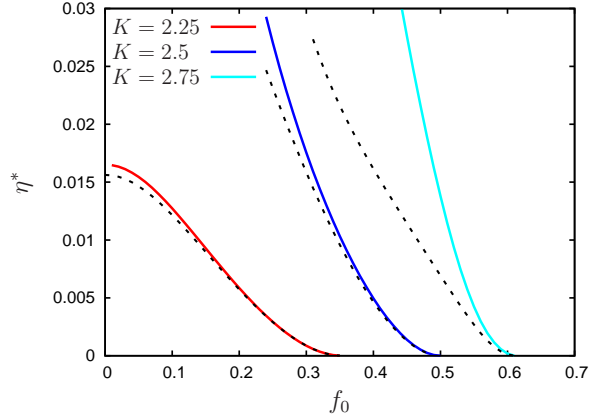}
\caption{EMP $\eta^*$ as obtained by maximizing $\Po$ wrt $s$,  as a function of $f_0$ for different values of the coupling constant $K$. The dashed lines correspond to the approximated expression (\ref{etass_approx}).}
\label{fig:opts}
\end{figure}

$iii)$ In order to optimize $\Po$, as given by eq.~(\ref{xx}), with respect to the quenched disorder  standard deviation $s$, one has to solve the equation  $\partial_{s} \Po=0$. This equation has no analytic solution $s^*$, but it can be solved numerically, in order to find the EMP for different values of $f_0$ and $K$, as shown in fig.~\ref{fig:opts}. Still one has the constraint $s^*>f_0$, in order for the force on the user to be negative.

One can also find an approximate expression for $s^*$  at the lowest order in $\Delta K$ and $s/T$ as follows.

By solving $\partial_{s} \Po=0$ for $K<K_c$, one easily finds that the maximum is given by $s_0^*=f_0$, and thus $\Po^*=0$.
For $K>K_c$, one can derive $\Po$ wrt $s$, and then expand the equations up to the second order in $(K-K_c)$ and $\Delta s=s-s_0^*$, solving for $s$, and plugging the value $s^*$ that maximize $\Po$ into the expression for $\eta$, one finds
\begin{eqnarray}
\eta^*&\simeq&\frac{(K-K_c)^2 T^4 \left(4 T^2+f_0^2\right)^2 \left(T^2+7 f_0^2\right)}{64 \left(T^2-2 f_0^2\right) \left(T^2+f_0^2\right) \left(T^4+3 T^2 f_0^2-f_0^4\right)^2}\nonumber \\
&\simeq& \frac{\Delta K^2} {T^2} \p{\frac 1 {4 } +\frac{5 f_0^2}{8 T^2}}, \label{etass_approx}
\end{eqnarray} 
where the last expression gives $\eta^*$ to the lowest order in $\Delta K$ and $f_0/T$. It is worth noting that the coefficients in the series expansions are the same as in eq.~(\ref{etass_app}).

Note that $\Delta K$ depends implicitly on $s^*$ through $K_c$, as given by eq.~(\ref{Kcs}), when one replaces $s$ with $s^*$.
Inspection of eq.~(\ref{xx}) suggests that the optimal quenched
disorder standard deviation $s^*$ increases as $f_0$ increases, but this in turn leads to a smaller value of $\Delta K$ for fixed $K$, as $K_c(s=s^*)$ also increases. Thus increasing $f_0$, and optimizing $\Po$ wrt $s$ drives the system towards smaller values of synchronization $\sigma$, resulting in a smaller $\eta^*$.
This is in agreement with the results reported in fig.~\ref{fig:opts}, showing that 
the EMP close to the dynamical phase transition, is enhanced by decreasing  the applied average force $f_0$.

\subsubsection{Gaussian distribution}

\begin{figure}[h]
\center
\psfrag{s0.3}[lc][lc][.7]{$s=0.3$}
\psfrag{s0.2}[lc][lc][.7]{$s=0.2$}
\psfrag{s0.15}[lc][lc][.7]{$s=0.15$}
\psfrag{s0.125}[lc][lc][.7]{$s=0.125$}
\psfrag{s0.1}[lc][lc][.7]{$s=0.1$}
\psfrag{dK}[ct][ct][1.]{$(K-K_c)/K_c$}
\psfrag{etas}[ct][ct][1.]{$\eta^*$}
\includegraphics[width=8cm]{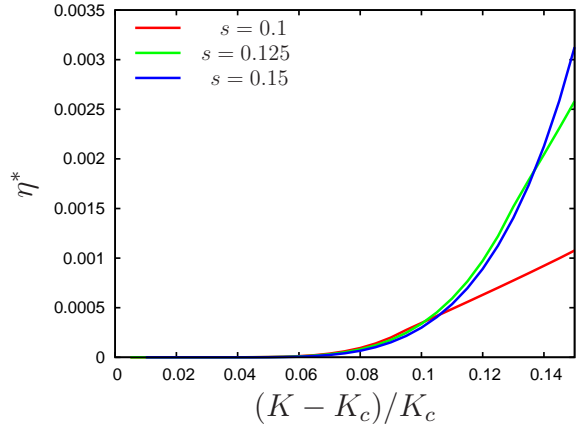}
\caption{EMP $\eta^*$, as obtained by maximizing $\Po$ wrt $f_0$,  as a function of $(K-K_c)/K_c$, for the Gaussian force distribution~(\ref{gauss:g}) for different values of the variance. }
\label{gauss:fig}
\end{figure}

We now consider the following distribution for the force
\begin{equation}
g(\omega)=\frac{\E^{-\frac{(\omega-f_0) ^2}{2 s^2}}}{\sqrt{2 \pi s^2} }.
\label{gauss:g}
\end{equation} 
From eq.~(\ref{kc:def}) we can calculate the critical coupling strength
\begin{equation}
K_c=2 \sqrt{\frac{2}{\pi }} s\frac{ \E^{-\frac{T^2}{2 s^2}}}{1-\erf\left(\frac{T}{\sqrt{2} s}\right)},
\end{equation} 
and from eqs.~(\ref{po:def}) and (\ref{pi:def}) we can calculate the output and the input power for $K<K_c$ which read
\begin{eqnarray}
{\Po}_{,0}&=&s f_0\frac{ e^{-\frac{f_0^2}{2 s^2}}}{\sqrt{2 \pi }}-\frac{1}{2} \left(s^2+f_0^2\right) \pq{1-\erf\left(\frac{f_0}{\sqrt{2} s}\right)}<0,\\
{\Pin}_{,0}&=&s f_0\frac{ e^{-\frac{f_0^2}{2 s^2}}}{\sqrt{2 \pi }}+\frac{1}{2} \left(s^2+f_0^2\right) \pq{1+\erf\left(\frac{f_0}{\sqrt{2} s}\right)}>0,
\end{eqnarray} 
with
\begin{eqnarray}
\lim_{s\to 0} {\Po}_{,0}&=&0, \qquad \lim_{\mf\to 0} {\Po}_{,0}=-\frac{s^2}{2},\\
\lim_{s\to 0} {\Pin}_{,0}&=&\mf^2, \qquad \lim_{\mf\to 0} {\Pin}_{,0}=\frac{s^2}{2}.
\end{eqnarray} 
No analytic result can be obtained for $\Po$, and $\Pin$ and thus for the EMP when $K>K_c$ for the Gaussian force distribution ~(\ref{gauss:g}), at variance with what has been done in the previous section. However, one can resort to numerical calculations, to integrate numerically eqs.~(\ref{po:def})--(\ref{pi:def}), find the optimal value of $\Po$ wrt to some parameter, and thus calculate the EMP. This procedure has been followed in order to calculate the EMP as obtained by maximizing $\Po$ wrt to the mean force $\mf$. The results are reported in fig.~\ref{gauss:fig}: while for small $\Delta K$ the EMP is larger the smaller the variance, for sufficiently large $\Delta K$
\begin{mycomments}
{\it ( numeric\_gaussian.nb)}
\end{mycomments}
 we find  the same tendency observed for the bimodal distribution, namely   $\eta^*$ as a function of $\Delta K$ is larger, the broader is the force distribution, and thus the degree of quenched disorder.

\subsection{Distribution of positive and negative forces on the same particle}\label{sec:both}
In this section, we consider the case where on the same particle, whose dynamics is described by eq.~(\ref{lang2}), two forces are applied, $f_{i,-}<0$ and $f_{i,+}>0$, distributed with two PDFs  $g_-(f_-)$ and $g_+(f_+)$. 
In this framework the output and input power read
\begin{eqnarray}
\Po&=&-\int_0^{+\infty}\D \op g_+(\op) \int^0_{-\infty} \D \om g_-(\om) \, \om \pq{\vp(\op+\om-f_0) + f_0}\label{poutx}, \\
\Pin&=&\int_0^{+\infty}\D \op g_+(\op) \int^0_{-\infty} \D \om g_-(\om) \, \op \pq{\vp(\op+\om-f_0) + f_0},
\end{eqnarray} 
respectively.

\subsubsection{No disorder}

We start our analysis by considering the trivial case where the same forces $\op$ and $\om$ are applied on all the particles, with $f_0=\op+\om>0$ and $\om<0$. 
So, the total force distribution reads $g(\omega)=\delta(\omega-f_0)$, and the delivered power (\ref{poutx}) becomes
\begin{equation}
\Po= -\om \cdot  f_0=-\om(\om+\op),
\end{equation} 
which is independent of the coupling $K$.  Thus by maximizing $\Po$ wrt $f_-$ one finds that the EMP is always $\eta^*=1/2$, as in the linear regime case \cite{Esposito2009,Golubeva2012,VandenBroeck2012,Seifert2011}.
\subsubsection{Bimodal negative force distribution}

\begin{mycomments}
{\it 
Different strategy: introduce $x=\omega_{1,-}+\omega_{2,-}$, and $y=\omega_{1,-}-\omega_{2,-}$, and optimize wrt $x$.

see two\_omega\_minus1.nb, bottom}
\end{mycomments}

In order to increase the complexity of our system, 
we consider here the following distributions
\begin{eqnarray}
g_+(f)&=& \delta (f-f_+),\\
g_-(f)&=& \frac 1 2 \pq{\delta (f-f_{1,-}) +\delta (f-f_{2,-})} \\
\end{eqnarray} 
so as the total force distribution reads
\begin{equation}
 g(f)=\frac 1 2 \pq{\delta(f-(\op+f_{1,-})) + \delta(f-(\op+f_{2,-}) }.
\end{equation} 

We introduce the following variables 
\begin{equation}
x=\frac {f_{1,-}+f_{2,-}}{2} <0, \qquad y=\frac{f_{1,-}-f_{2,-}}{2}.
\end{equation} 
Given the expression for the average force, we obtain a condition on $x$:
\begin{equation}
f_0=\op+\frac{f_{1,-}+f_{2,-}} {2} =\op+ x >0 \quad \Rightarrow  \quad - \op < x <0.
\end{equation} 
The delivered power eq.~(\ref{poutx}) thus becomes
\begin{equation}
\Po=-  \pq{x (x+ \op) + y v_\theta\p{ y }},
\label{pox1}
\end{equation} 
while the input power reads
\begin{equation}
\Pin=  \op (x+ \op).
\end{equation} 
It turns out that with this choice of the force distributions the order parameter $\sigma$ and thus $v_\theta$ depends only on the negative forces, indeed we have
\begin{eqnarray}
 g_0(\omega)&=&g(\omega+f_0)=\frac 1 2 \pq{\delta(\omega-(f_{1,-}-f_{2,-})/2) + \delta(\omega-(f_{2,-}-f_{1,-})/2) }\nonumber \\
&=& \frac 1 2 \pq{\delta(\omega-y) + \delta(\omega+y) },
\end{eqnarray} 
and $g_0(\omega)$ is the function appearing in the self consistent eq.~(\ref{eqsig}).

By deriving $\Po$ with respect to the disorder degree parameter $y$, we obtain $\partial_y \Po=-(v_\theta\p{ y }+y v_\theta'\p{ y })<0$, where we have assumed $y>0$ without loss of generality, i.e. $\Po$ decreases monotonically with $y$. Thus the optimal $\Po$ is trivially obtained for $y=0$ which corresponds to the case with no disorder, with  $\eta^*=f_-/f_+$.

\begin{mycomments}
{\it
Note that there is a condition for $\Po>0$, indeed
$\Po=-  \pq{x (x+ \op) + y v_\theta\p{ y }}\ge -x (x+\op)-y^2=-x f_0-y^2$
}

\end{mycomments}

We now optimize $\Po$ eq.~(\ref{pox1}) wrt the variable $x$, which is equal to the mean negative force, and find 
\begin{equation}
x^*=-\frac {\op}{2},
\end{equation} 
and thus
\begin{equation}
\Po^*=\frac 1 4 \op^2- y  v_\theta\p{ y }, \qquad \Pin^*=\frac {\op^2}{2},
\label{posy}
\end{equation} 
and finally the EMP reads
\begin{equation}
\eta^*= \frac {\op^2-4 y  v_\theta\p{ y}}{ 2 \op^2}.
\end{equation} 
We notice that while for $K<K_c$ one finds $v_\theta(y)=y$, and therefore in the uncoupled regime $\eta^*<1/2$, for $K>K_c$ the velocity $v_\theta(y)$ is a decreasing function of $K$, thus for a fixed $y$ the coupling reduces the spread around the mean velocity $f_0$, and thus in the limit of large $K$ one recovers the single particle EMP $\eta^*=1/2$.
\begin{mycomments}
{\it 
 ($v_\theta(y)$ is an odd function of $y$, with $v_\theta(y)>0$ if $y>0$, so the product $yv_\theta(y)>0$).}
\end{mycomments}
Similarly, $\Po^*$ (\ref{posy}) is an increasing function of $K$, because of the decreasing behaviour of  $v_\theta(y)$.
By expanding $v_\theta$ in powers of $K$, and noticing that the order parameter $\sigma$ is given by eq.~(\ref{s_s}) , with the substitution $s\rightarrow y$, 
we obtain
\begin{equation}
\eta^*\simeq \frac {\op^2- 4 y^2}{ 2 \op^2}+ \frac{4 y^2}{T \op^2}\Delta K,
\label{etasx}
\end{equation} 
to the lowest order in $\Delta K$ and $y$.
We notice that the quantity $y^2$ is equal to the variance of the negative forces' distribution, thus the introduction of disorder in the force distribution, on the one hand reduces the EMP  for an uncoupled system ($K<K_c$, first term on the rhs of eq.~(\ref{etasx})), on the other hand it increases the slope of the EMP above the critical coupling.

\begin{figure}[h]
\center
\psfrag{0.25}[lc][lc][.7]{$y=0.25$}
\psfrag{0.30}[lc][lc][.7]{$y=0.30$}
\psfrag{0.35}[lc][lc][.7]{$y=0.35$}
\psfrag{0.5}[lc][lc][.7]{$y=0.5$}
\psfrag{0.6}[lc][lc][.7]{$y=0.6$}
\psfrag{0.7}[lc][lc][.7]{$y=0.7$}
\psfrag{K}[ct][ct][1.]{$K$}
\psfrag{etas}[ct][ct][1.]{$\eta^*$}
\includegraphics[width=8cm]{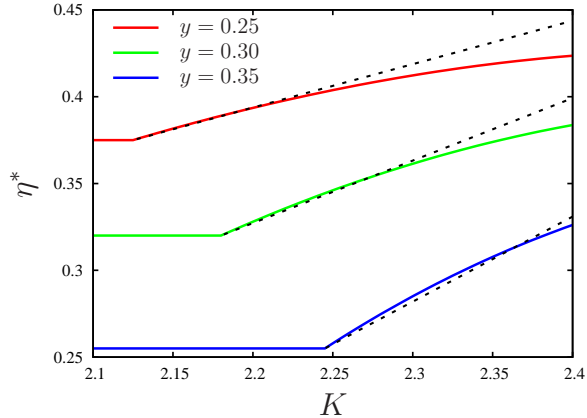}
\caption{EMP as a function of the coupling strength $K$ as obtained by maximizing $\Po$ wrt the average negative force $x$, with $T=1$, and $\op=1$. The
dashed lines correspond to the approximate expression (\ref{etasx}).}
\label{fig321}
\end{figure}

In fig.~\ref{fig321} the EMP $\eta^*$ is plotted as a function of $K$ for different values of $y$.

\begin{mycomments}
{\it 
fig.~(\ref{fig321}) done with $\sigma s4$ in mathematica

$\Po$ eq.~(\ref{pox1}) has not a maximum wrt to $y$. Indeed $\partial_y \Po=v_\theta(y)+ y v'_\theta(y)\ge 0$, as both the terms in the sum are positive for $y>0$.}
\end{mycomments}
\subsubsection{General case: optimization}
We now consider the case where the system exhibits a  distribution of both positive and negative forces $g_-(f_-)$ and $g_+(f_+)$.
The distribution of the total forces on each particle thus reads
\begin{equation}
g(f)=\int \D f_-\, \D f_+\,  g_-(f_-)g_+(f_+) \delta(f-(f_++f_-)),
\end{equation} 
with 
\begin{equation}
f_0=\int \D f_-\, g_-(f_-)f_-+\int \D f_+\,  g_+(f_+)=\bar f_-+\bar f_+,
\end{equation} 
and
\begin{equation}
g_0(\omega)=g(\omega+f_0)=\int \D f_-\, \D f_+\,  g_-(f_-)g_+(f_+) \delta(\omega-(y_++y_-)),
\end{equation} 
where $y_\pm=f_\pm-\bar f_\pm$.
The distribution $g_0(\omega)$ is symmetric around $\omega=0$ if both the distributions $g_\pm(f_\pm)$ are symmetric around their respective average value $\bar f_\pm$, a symmetry that we assume in the following.

We have thus
\begin{eqnarray}
\Po&=&-\int_0^{+\infty}\D y_+ g_+(\op) \int^0_{-\infty} \D y_- g_-(\om) \, \om \pq{\vp(\op+\om-f_0) + f_0}\nonumber\\
&=&-\bar f_-\p{\bar f_++\bar f_-}-\int_{-f_+}^{+\infty}\D y_+ g_+(y_+) \int^{-f_-}_{-\infty} \D y_- g_-(f_-) \, y_-\vp(y_++y_-), \nonumber
\end{eqnarray} 
and similarly
\begin{equation*} 
\Pin=\bar f_+\p{\bar f_++\bar f_-}+\int_{-f_+}^{+\infty}\D y_+ g_+(y_+) \int^{-f_-}_{-\infty} \D y_- g_-(y_-) \, y_+\vp(y_++y_-),
\end{equation*} 
where in the last equality we have used the above mentioned symmetry of $g_\pm(f_\pm)$.
Thus, if we want to optimize $\Po$ wrt to $\bar f_-$, we obtain
\begin{equation}
\bar f_-^*=-\frac{f_+ }{2}
\end{equation} 
and
\begin{eqnarray}
\Po^*&=&\frac{\bar f_+^2 }{4}-\int_{-f_+}^{+\infty}\D y_+ g_+(y_+) \int^{-f_-}_{-\infty} \D y_- g_-(y_-) \, y_- v_\theta(y_++y_-),\\
\Pin^*&=&\frac{\bar f_+^2 }{2}+\int_{-f_+}^{+\infty}\D y_+ g_+(y_+) \int^{-f_-}_{-\infty} \D y_- g_-(y_-) \, y_+ v_\theta(y_++y_-).
\end{eqnarray} 
We can thus evaluate the optimal delivered power below the critical coupling, and for very large coupling constant

\begin{equation}
\label{cases}
\Po^*=\cases{\frac{\bar f_+^2 }{4}- \average{y_-^2},  &for $K \le K_c$\\
\frac{\bar f_+^2 }{4},  &for $K\gg K_c$\\
}
\end{equation}
Close to the critical $K$ up to the second order in $\sigma$, by using eq.~(\ref{v_exp}), we find
\begin{eqnarray}
\Po^*&=&\frac{\bar f_+^2 }{4}- \average{y_-^2}+\int \D y_+\D y_- g_+(y_+) g_-(y_-)  \frac{y_-(y_-+y_+)K^2 \sigma^2}{2 \pq{T^2+(y_-+y_+)^2}}\nonumber \\&\simeq&\frac{\bar f_+^2 }{4}+ \average{y_-^2}\p{\frac{2 \Delta K}{T}-1},\\
\Pin^*&=&\frac{\bar f_+^2 }{2}+ \average{y_+^2}+\int \D y_+\D y_- g_+(y_+) g_-(y_-)  \frac{y_+(y_-+y_+)K^2 \sigma^2}{2 \pq{T^2+(y_-+y_+)^2}}\nonumber \\ &\simeq&\frac{\bar f_+^2 }{2}+ \average{y_+^2}\p{\frac{2 \Delta K}{T}+1},
\end{eqnarray} 
to the lowest order in $\Delta K/T$.
We obtain finally the EMP
\begin{equation}
\eta^*=\frac 1 2 \frac{\bar f_+^2-4 \average{y_-^2}}{\bar f_+^2+2 \average{y_+^2}}+ \frac{2\Delta K}{ T} \frac{\bar f_+^2 ( 2 \average{y_-^2}- \average{y_+^2})+8 \average{y_-^2} \average{y_+^2}}{\p{\bar f_+^2+2 \average{y_+^2}}^2}.
\label{last:eq}
\end{equation} 
inspection of this last equation suggests that the maximal slope of $\eta^*$ as a function of $\Delta K$ is obtained for $ \average{y_+^2}=0$.
As far as the variance of the negative forces is concerned, we find a similar scenario as in the previous section: while $\average{y_-^2}$ reduces the EMP for the uncoupled system, above the critical coupling the EMP increases faster the larger is $\average{y_-^2}$.
\section{Conclusions}
In the present paper we have investigated the thermodynamic properties of a model of microscopic oscillators, subject to thermodynamic forces.
We considered the effect of the disorder on the delivered and injected power and on the  EMP, and discussed the critical behavior of such  quantities for different force distributions.

We considered two forces distribution types, one that resembles the macroscopic power grids, and one that resembles a system of interacting autonomous motors.

For the first type of force distribution we find that, at fixed coupling strength, a larger degree of disorder leads to an increase in the EMP, at least close to the critical point.

For the second type of force distribution we find that while the disorder reduces both the optimal $\Po$ and the EMP below the critical coupling, above the critical point the EMP rate as a function of $\Delta K$ increases as the degree of disorder increases.

Thus, ideally the system with the optimal thermodynamic performances is characterized by a strong coupling ($K\to \infty$) or absence of force disorder.
However, in a real system one may have to deal with a finite coupling strength and an intrinsic non--vanishing degree of disorder in the force distribution.
The results contained in this paper characterize the thermodynamic properties of such systems.
 
While 
we were able to calculate the expansion of the energy rates and of the EMP close to the critical point, we found that the EMP does not exhibit any universal behaviour, at variance with the single device case. On the contrary, the results contained in this paper depends strongly on the details of the force distribution.
For example, in eq.~(\ref{last:eq}), one recovers the EMP value in the linear regime ($\eta^*=1/2$) only when the disorder vanishes.

One of the limitations of the present model is that it exhibits an "all-to-all'' coupling, while in a real system the interaction can depend on the distance between the network nodes and on the network topology, thus one should replace the interaction strength $K$ in eq.~(\ref{lang1}) with an interaction matrix $K_{ij}$. The thermodynamic properties of this extended model are certainly worth to investigate.

Furthermore, the values of the EMP reported in figs.~\ref{fig:either}, \ref{fig:opts}, \ref{gauss:fig} are quite small, thus the characterization of the response of the network injected and delivered power, or of its efficiency to a change in the network  topology is certainly  worthy of future investigation. For example, one may want to find the connectivity matrix between the different nodes that optimize the relevant thermodynamic quantities. 
\section*{Acknowledgments}
This work was supported by  the Danish
Council for Independent Research, and the COST Action MP1209 ``Thermodynamics in the Quantum Regime.

\section*{References}
\bibliography{bibliography}

\end{document}